# Microwave probe sensing location for Venturi-based real-time multiphase flowmeter

Mengke ZHAN[a,b], Cheng-Gang XIE[b], Jian-Jun SHU[a,*]

[a] *School of Mechanical & Aerospace Engineering, Nanyang Technological University, 50 Nanyang Avenue, Singapore 639798*
[b] *Schlumberger Oilfield (Singapore) Pte Ltd, Singapore Well Testing Center, 1 Benoi Crescent, Singapore 629986*

ABSTRACT

Real-time in-line interpretation of liquid properties is important for multiphase flow measurements; for example, many phase fraction measurement principles have a dependency on characteristics of salinity. Therefore, it would be desirable to have a sensor, such as a microwave sensor, which can continuously measure the salinity of a flow. In addition to salinity measurement, the microwave sensor can also measure water fraction, which is required for a multiphase flowmeter based on single-energy gamma-ray attenuation; however, choosing a suitable probe sensing location for a microwave salinity sensor in a multiphase flowmeter can be challenging, as the sensor needs to be located at near-wall liquid-rich region to accommodate a wide range of flow conditions. Currently, a microwave sensor is installed in the lower area of a horizontal blind-tee inlet spool of a multiphase flowmeter for salinity measurement. Integrating the microwave sensor into the vertically mounted multiphase flowmeter can reduce the flowmeter (carbon) footprint and manufacturing costs and can improve water-to-liquid ratio measurement due to faster local oil-water mixing. The associated challenge is that the sensor needs to be located at near-wall liquid-rich region to accommodate a wide range of flow conditions, including high gas-volume-fraction flows, where the near-wall liquid layer present in the vertical pipe is usually very thin. In this study, computational fluid dynamics modeling is used to evaluate the suitability of four different sensing locations along the vertical cross-section of a Venturi-based multiphase flowmeter based on near-wall liquid-richness. The results show that the Venturi inlet is the most suitable location for microwave sensor measurement, compared to the mid-convergence section, the mid-divergence section, and the Venturi outlet for a range of inlet liquid-volume-fractions. The findings have been validated by experimental microwave sensor measurements in a multiphase flow loop facility.

* Corresponding author.
 *E-mail address:* mjjshu@ntu.edu.sg (Jian-Jun SHU).

**Keywords**: Multiphase flowmeter; salinity measurement; water-to-liquid ratio measurement; Venturi tube; Eulerian–Eulerian modeling; microwave sensing.

# 1 Introduction

A multiphase flowmeter (MPFM) has been used in the upstream oil and gas industry for continuous, in-line, real-time flow measurement of oil-gas-water without the need for fluid separation (bin Razali *et al.*, 2021). An MPFM consists of phase fraction (holdup) and velocity (or flow rate) measurements. In the phase fraction measurement technique, single-energy gamma-ray attenuation is commonly used to measure the chord-averaged phase fraction of gases and liquids with sufficiently different densities (Sætre *et al.*, 2010). Through dual-energy gamma-ray attenuation, the phase fraction of gas, oil, and water can be measured to calculate the water-to-liquid ratio (WLR) (Chazel *et al.*, 2014). For an MPFM based on single-energy gamma-ray attenuation, only the fraction of gas or liquid can be measured; an additional measurement of the flowing mixture conductivity and/or permittivity can be used to determine the water fraction (and hence the WLR). In addition to the dependence on temperature, water conductivity and permittivity, which are typically required when calculating the water fraction for a water-continuous (high WLR) flow, are also dependent on salinity. Dual-energy gamma attenuation has a dependency on salinity change, since the photoelectric absorption that dominates at low energy, is highly dependent on atomic number, which varies in different concentration of brine solution (Johansen & Jackson, 2000). Therefore, it is valuable to track on-line changes in water properties over time through salinity measurement (Fiore *et al.*, 2020). While flow sampling can be used to measure salinity, in certain circumstances, sampling may fail to track rapid changes in the salinity of produced water, due to, for example, sharp salinity gradient in the formation (McCoy *et al.*, 1997) or water injection during enhanced oil recovery (Johansen & Jackson, 2000).

Microwave sensors operating in the GHz-range are suitable for tracking liquid properties such as the salinity, oil and water fractions of multiphase flow. A schematic diagram explaining the measurement principle of a nonintrusive open-ended coaxial microwave sensor is shown in Fig. 1. To prevent any intrusion of the sensor from disturbing the flow, the microwave sensor is mounted flush with the inner pipe wall of an MPFM. The incident microwave interacts with the multiphase flow with a depth of investigation of a few millimeters. The portion of the signal that is not absorbed by the flow is reflected. The amount of absorption depends on the electromagnetic properties of the fluid composition. For example, a water-rich mixture has more absorption than an oil-rich mixture. Hence, the electrical conductivity and dielectric permittivity of the fluid mixture in contact with the sensor can be inferred from the amplitude-attenuation and phase-shift of the reflected signal relative to the incident signal. Due to the sharp contrast of the permittivity and conductivity of water relative to oil and gas hydrocarbons, the WLR can be interpreted from the measured mixture permittivity and mixture conductivity (Fiore *et al.*, 2020). Salinity can be obtained from the ratio of the mixture conductivity to the mixture permittivity at a given fluid temperature (Xie *et al.*, 2004).

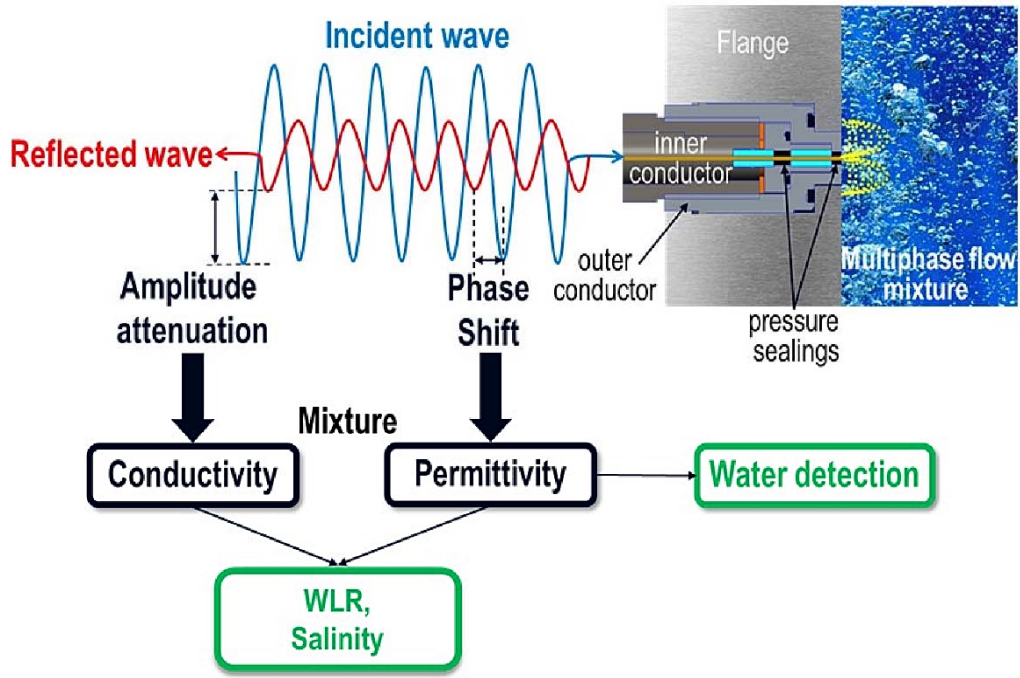

**Fig. 1.** Schematic diagram of open-ended coaxial microwave sensor and signal interpretation (Fiore *et al*., 2020)

Currently, the microwave sensor is installed in the lower area of a horizontal blind-tee spool upstream of a vertically mounted Venturi-based MPFM for salinity measurement (Xie *et al*., 2013; Fiore *et al*., 2020). Even though the local flow in the lower area of the horizontal blind-tee is liquid-rich, it is desirable to make the flowmeter installation more compact by integrating the microwave sensor with the vertically mounted MPFM. This can reduce the manufacturing cost and (carbon) footprint of the flowmeter. In addition, measurement of the flowing mixture conductivity and/or permittivity can be used to determine the liquid WLR; however, water and oil in the horizontal blind-tee may separate due to their density differences and low local flow rates. Hence, it is desirable to position the microwave sensor in a region with better oil-water mixing, thus with a more representative WLR, such as a suitable near-wall region for the vertical Venturi MPFM; however, since the investigation depth of an open-ended coaxial microwave sensor is only a few millimeters, so the sensing location along the vertical Venturi must be carefully studied, given that the liquid layer near the wall can be thin with potential gas entrainment in a highly turbulent flow. To select the suitable sensing location of the microwave sensor for a vertically mounted Venturi-based MPFM, a computational fluid dynamics (CFD) model is used as the first approach to assess near-wall liquid-richness (LR) at several locations along the Venturi profile. These findings are subsequently validated by microwave sensor data obtained from flow loop experiments.

## 2 Mathematical model

### 2.1 Eulerian–Eulerian multiphase model

Many CFD studies have been performed on gas-liquid flow. Some of the most common modeling approaches include the Eulerian–Eulerian model (Shu & Wilks, 1995; Yamoah *et al.*, 2015; Zhang *et al.*, 2019; Acharya & Casimiro, 2020), the volume of fluid model (Frankiewicz *et al.*, 2001; Shu, 2003b; Laleh *et al.*, 2011; López *et al.*, 2016), and the mixture model (Hallanger *et al.*, 1996; Shu *et al.*, 1997; Shu, 2003a; Shang *et al.*, 2015). In this study, the Eulerian–Eulerian model is used as a modeling approach to solve the ensemble-averaged mass and momentum transport equations for dispersed gases and continuous liquids, allowing the details of phase interaction forces to be modeled. Some assumptions are made to simplify the Eulerian–Eulerian governing equations.

First, a steady framework is used instead of a transient one, because (i) the flow loop experiment used for validation has steady inlet condition; (ii) the flow rate output is computed at a frequency of 1 Hz for gamma-ray full-spectrum deconvolution (Chazel *et al.*, 2014), which can be considered as the "average flow rate" over a second duration; (iii) in steady flow, steady-state simulation produces the equivalent of time-averaged transient results. To show that the flow reaches a steady-state, the time-series data of the water-nitrogen flow rates for the Venturi throat chord-average liquid fraction $\alpha_l$ and the Venturi differential pressure $\Delta P$ computed at 1 Hz every 10 seconds in the duration of 800 seconds are shown in Fig. 2a and Fig. 2b, respectively.

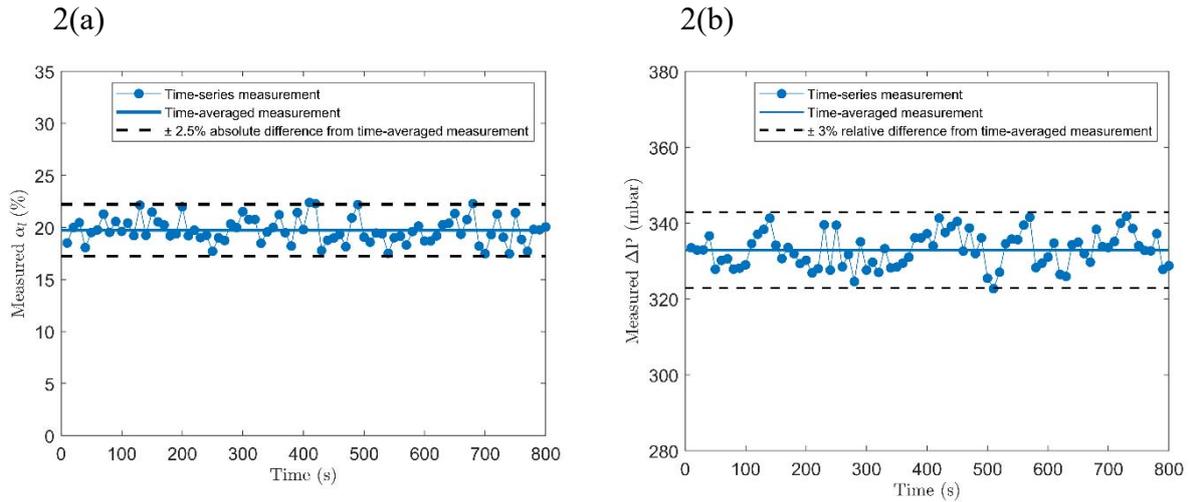

**Fig. 2.** Variation of (a) liquid fraction $\alpha_l$ against time (with coefficient of variation =1.56%) and (b) $\Delta P$ against time (with coefficient of variation =1.38%). The water-nitrogen test point has inlet liquid-volume-fraction ~ 16% and flow inlet velocity ~ 5 m/s.

From Fig. 2, the variation of both measured throat chord-average liquid fraction $\alpha_l$ and the Venturi $\Delta P$ is insignificant over time with the respective coefficient of variation or standard deviation-to-mean ratio less than 2%. The variation of chord-average liquid is mostly within ±2.5% absolute from the time-averaged measurement, while that of $\Delta P$ is mostly within ±3% relative from the time-averaged measurement. Therefore, the steady-state solver is used to study the "time-averaged" behavior of gas-liquid flow because it is computationally more economical.

The second assumption is that the gas-liquid flow is considered incompressible if it satisfies the criterion, $\frac{\Delta P}{P} < 3\%$. In the experiments, nitrogen gas is used as the gas phase. According to the Aungier–Redlich–Kwong equation of state (Aungier, 1995), the nitrogen density difference between the flow inlet and the throat with respect to the throat density is less than 3%, under the experimental condition at pressure $P \sim 20$ and temperature $T \sim 30°C$.

A further assumption is that there is no mass transfer between phases; there is no mass source term or external force, apart from gravitational force. The simplified Eulerian–Eulerian governing equations are shown in Equations (1)–(4).

$$\sum_{q=1}^{n} \alpha_q = 1, \qquad (1)$$

$$\nabla \bullet \left( \alpha_q \vec{v}_q \right) = 0 , \quad (2)$$

$$\rho_q \nabla \bullet \left( \alpha_q \vec{v}_q^2 \right) = -\alpha_q \nabla P + \nabla \bullet \bar{\bar{\tau}}_q + \alpha_q \rho_q \vec{g} + \sum_{p=1}^{n} \vec{F}_{pq} , \quad (3)$$

$$\bar{\bar{\tau}}_q = \alpha_q \mu_q \left( \nabla \vec{v}_q + \nabla \vec{v}_q^T - \frac{2}{3} \nabla \bullet \vec{v}_q \bar{\bar{I}} \right) , \quad (4)$$

where $\alpha_q$, $\vec{v}_q$, $\rho_q$ and $\bar{\bar{\tau}}_q$ are the phase fraction (holdup), velocity, density and stress tensor of phase $q$, respectively, and $P$, $\vec{g}$, $\vec{F}_{pq}$, $\bar{\bar{I}}$ and $\mu_q$ are the pressure shared by all the phases, gravitational constant, interphase force acting on phase $q$, identity tensor and shear viscosity of phase $q$, respectively. For gas-liquid two-phase flow, the number of phases is $n = 2$.

The turbulence model of multiphase flow includes dispersed formulation and mixture formulation. Dispersion formulations are used when the dispersed phase is dilute. Mixture formulation assumes that all phases share the same turbulence flow field. The mixture properties, such as the mixture density, viscosity and velocity, are used in the mixture formulation (Cokljat *et al.*, 2006). A mixture shear stress transport $k-\omega$ turbulence model (Menter, 1994) is used in the study.

*2.2 Interphase forces*

There are diverse types of interphase forces under $\vec{F}_{pq}$ in the phase momentum equation, such as virtual mass, drag, lift, wall lubrication and turbulent dispersion forces. This study identifies that drag force $\vec{F}_D$, lift force $\vec{F}_L$, wall lubrication force $\vec{F}_{WL}$, and turbulent dispersion force $\vec{F}_{TD}$ are the dominant interphase forces, when modeling water-nitrogen and oil-nitrogen flows with negligible compressibility, as shown in Equation (5), where $\vec{F}_{qp}$ is the interphase force acting on the phase $p$. The contribution of the virtual mass force to the simulation results in terms of phase holdup and differential pressure is insignificant and neglected in the modeling. The insignificance of virtual mass forces has also been reported in the literature (Kriebitzsch & Rzehak, 2016; Lote *et al.*, 2018; Colombo & Fairweather, 2019; Kim & Park, 2019).

$$\vec{F}_{pq} = -\vec{F}_{qp} = \vec{F}_D + \vec{F}_L + \vec{F}_{WL} + \vec{F}_{TD}. \tag{5}$$

The drag force $\vec{F}_D$ includes skin drag, where the dispersed phase is subjected to viscous force, and a drag is created where the pressure difference arises from the flow separation that occurs depending on the shape of the dispersed phase (bubbles or droplets) (Lote *et al.*, 2018). The drag force $\vec{F}_D$ is formulated as Equation (6).

$$\vec{F}_D = \frac{3}{4}\frac{C_D}{d_p}\alpha_p \rho_q \left|\vec{v}_p - \vec{v}_q\right|\left(\vec{v}_p - \vec{v}_q\right), \tag{6}$$

where $d_p$ and $C_D$ stand for the Sauter mean diameter (Sauter, 1926) and drag coefficient, respectively. The Tomiyama's drag model (Tomiyama *et al.*, 1998) is used to simulate the drag coefficient $C_D$ in the study. In the Tomiyama's model, the Eötvös number, $E_o$ (based on the Sauter mean diameter of the gas phase $d_p$), and the particle Reynolds number, $R_{eP}$, are used to determine the governing effect and resultant drag force of different bubble shapes.

The lift force $\vec{F}_L$ is a transverse force experienced when fluid particles move in a shear flow, and more generally in a rotational flow (Auton, 1987). The lift force $\vec{F}_L$ is formulated as Equation (7).

$$\vec{F}_L = -C_L \alpha_p \rho_q \left(\vec{v}_p - \vec{v}_q\right) \times \left(\nabla \times \vec{v}_q\right), \tag{7}$$

where $C_L$ is the lift coefficient. The Tomiyama's lift model (Tomiyama *et al.*, 2002) is used to model the lift coefficient $C_L$ under study.

The wall lubrication force $\vec{F}_{WL}$ is caused by the pressure difference resulting from the velocity difference of the liquid between the bubble and the wall and the liquid away from the bubble toward the pipe center (Antal *et al.*, 1991). The wall lubrication force $\vec{F}_{WL}$ is formulated as Equation (8).

$$\vec{F}_{WL} = -C_{WL}\alpha_p\rho_q\left|\vec{v}_p - \vec{v}_q\right|^2 \hat{n}_W, \qquad (8)$$

where $C_{WL}$ and $\hat{n}_W$ are the wall lubrication force coefficient and the normal component pointing away from the wall, respectively. The Frank wall lubrication model (Frank *et al.*, 2008) is used to model the wall lubrication force coefficient $C_{WL}$; this model removes the pipe-diameter dependency from the earlier model (Tomiyama, 1998) and is applicable for a wide range of geometries.

In a turbulent dispersed multiphase flow, the turbulent continuous phase can interact with the dispersed phase. The dispersed phase can be trapped in the continuous phase turbulent eddy and carried from the regions of high concentration to the regions of low concentration. The turbulent dispersion force $\vec{F}_{TD}$ plays an important role in the lateral phase distribution. The turbulent dispersion model developed by Burns *et al.* (2004) based on the Favre average of the interphase drag force is used in the study. The turbulent dispersion force $\vec{F}_{TD}$ is formulated as Equation (9).

$$\vec{F}_{TD} = C_{TD}\frac{3}{4}\frac{C_D\rho_q\nu_t}{d_b\sigma_{TD}\left(1-\alpha_p\right)}\left|\vec{v}_p - \vec{v}_q\right|\nabla\alpha_p, \qquad (9)$$

where $C_{TD}$, $\sigma_{TD}$ and $\nu_t$ are the turbulent Schmidt number for continuous phase volume fraction, turbulent dispersion coefficient and turbulent kinematic viscosity, respectively.

Lift, drag, wall lubrication, and turbulent dispersion force depend on the size of the Sauter mean diameter. In particular, there is a threshold diameter above which the lift changes direction.

This effect describes the phenomenon of small bubbles tending to move toward the pipe wall and large bubbles moving toward the center. The size of the Sauter mean diameter is obtained through the sensitivity study on a single test point of water-nitrogen and oil-nitrogen flows. The value that best predicts the measured Venturi differential pressure is selected from the sensitivity study. The Sauter mean diameter size obtained from the sensitivity study is then used in the simulation of other respective test points under a wide range of flow conditions.

## 3 Simulation setup

### 3.1 Geometry

The geometry of the computational model is shown in Fig. 3. The flow enters the horizontal section from the bottom left, passes through the horizontal-to-vertical blind-tee, and leaves from the horizontal section from the top right. The locations of the Venturi inlet (VI) (pipe-diameter, $D$, upstream of convergence section), the mid-convergence section (MC), the mid-divergence section (MD) and the Venturi outlet (VO) (pipe-diameter $D$ downstream end of divergence section) are shown in Fig. 3.

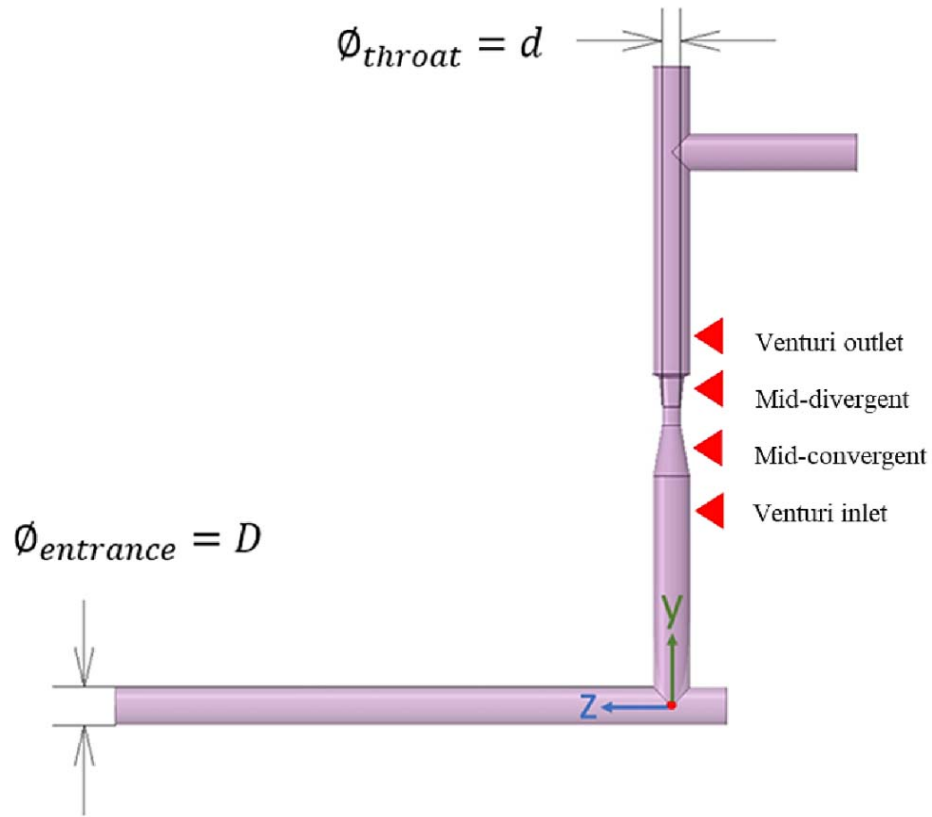

**Fig. 3.** Geometry of three-dimensional computational model, positive $x$ direction is pointing out of plane.

*3.2 Meshing*

Polyhedral cells are used for meshing to achieve the best performance in accuracy and computational efficiency among other cell shapes such as tetrahedral cells (Milovan & Stephen, 2004). The factors of global size (for local refinement) of the Venturi's convergence, divergence, and throat sections are $0.5$, $0.5$ and $0.25$, respectively. Ten inflation layers with a transition ratio $0.272$ and a growth rate $1.1$ are applied at the near-wall region to capture the boundary layer development. Cross-sectional and front views of the mesh showing a local refinement are shown in Fig. 4. Mesh-independent study is performed only by reducing the global size. Therefore, the mesh is scaled down. The simulated gamma-ray beam equivalents $\alpha_l$ and $\Delta P$ are used as the criterion for mesh-independence study as they are used for CFD model validation. An example of a mesh-independence study is shown in Fig. 5. The mesh with a cell number of 2.06 million is selected for the study. The cell number is further increased to 2.69 million (30.58% increase), with only 0.08% absolute change in $\alpha_l$ and 0.4% relative change in $\Delta P$.

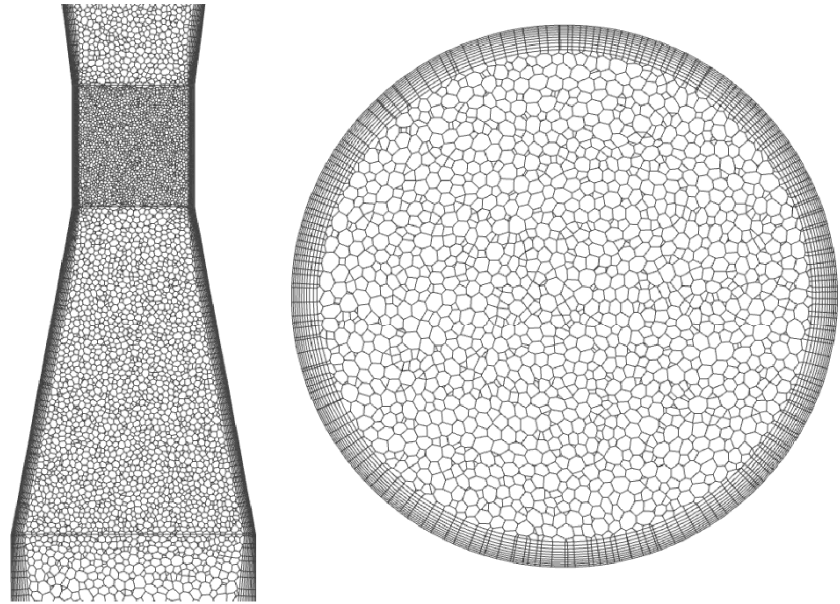

**Fig. 4.** Front (left) and cross-sectional (right) views of mesh

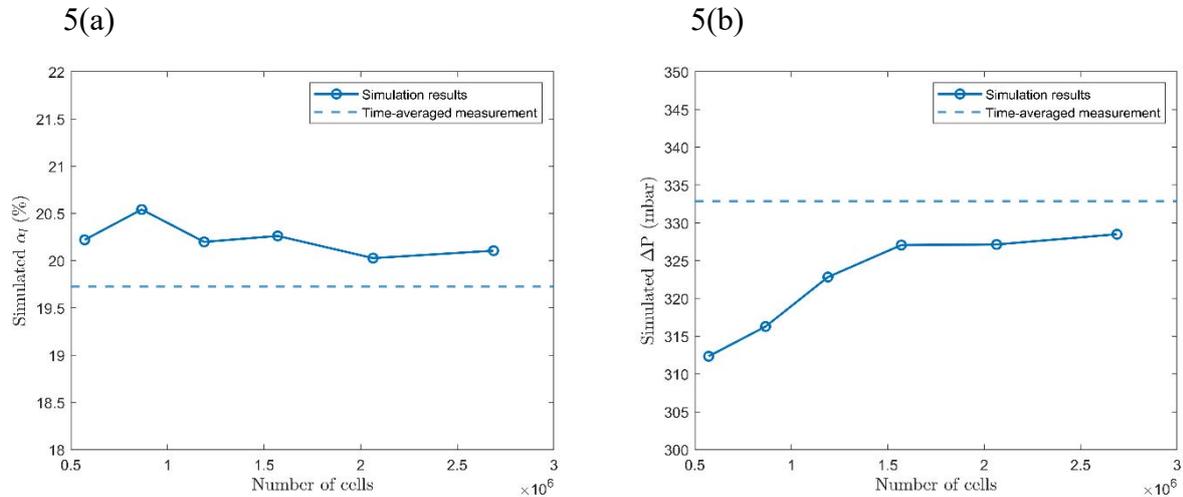

**Fig. 5.** An example of mesh-independence study for a water-nitrogen test point: (a) simulated chord-averaged liquid fraction $\alpha_l$ against number of cells and (b) simulated $\Delta P$ against number of cells. The water-nitrogen test point has inlet liquid-volume-fraction ~ 16% and flow inlet velocity ~ 5 m/s.

The simulation input includes boundary conditions, material properties and numerical schemes. Homogeneous phase velocity and fraction, turbulence intensity (~ 4% estimated), and pipe hydraulic diameter are specified at inlet conditions. Zero-gauge pressure is set at the outlet (assuming the flow has a negligible compressibility effect on the selected flow conditions). Material properties include the line-condition density and viscosity of the fluid. The surface tension of the liquid is also required to calculate the interaction force. The phase-coupled SIMPLE (Semi-Implicit Method for Pressure Linked Equations) algorithm and the first-order upwind scheme are used for spatial discretization. Simulation is performed on the CFD software Ansys Fluent 2021 R1.

## 4 Result and discussion

### 4.1 Quantitative validation of chord-averaged liquid fraction and venturi $\Delta P$

Before using a CFD model to investigate the suitability of microwave probe measurement at different sensing locations along a vertical Venturi, simulation results are validated against experimental measurements. Validation is performed based on the time-averaged Venturi $\Delta P$ measured between the throat midplane and the pipe-diameter upstream of the convergence section and the diametrically chord-averaged liquid fraction $\alpha_l$ measured by the gamma-ray sensor. CFD simulations are performed on five test points of water-nitrogen and oil-nitrogen flows from the experimental multiphase flow loop, covering a wide-range of flow inlet liquid-volume-fraction (LVF) from ~16% to ~74%. The test points have a range of flow inlet homogeneous velocities from 1.46 m/s to 6.07 m/s. Simulated values for the measurements of chord-averaged liquid fraction $\alpha_l$ and the $\Delta P$ in the throat are shown in Fig. 6a and Fig. 6b, respectively.

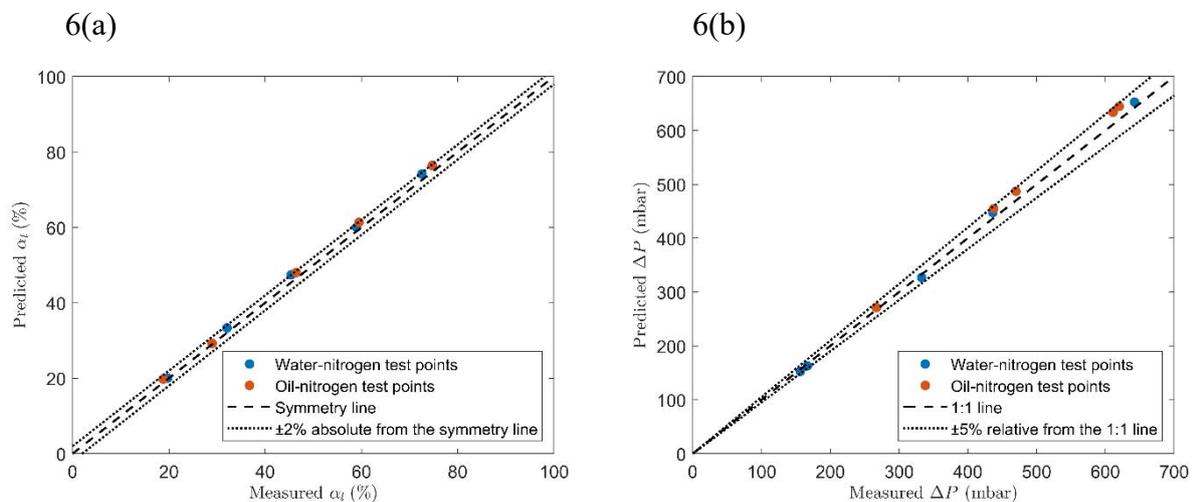

**Fig. 6.** Computational fluid dynamics prediction of (a) chord-averaged liquid fraction $\alpha_l$ at throat against gamma-ray measurement and (b) Venturi $\Delta P$ against measurement

For the absolute difference between the predicted and measured chord-averaged liquid fraction $\alpha_l$, all ten test points have a difference $\leq 2\%$. In terms of the Venturi $\Delta P$, all ten test points have a relative difference $< 5\%$. Overall, the predicted results of the chord-averaged liquid fraction and differential pressure at the throat section are close to the measured values.

*4.2  Evaluation of near-wall liquid-richness*

In this study, the near-wall LR is evaluated using two quantities: (i) the one-dimensional near-wall average liquid fraction $\bar{\alpha}_{lf}$ over the sensing depth of the microwave sensor probe, as indicated in Equation (10) and (ii) the near-wall LR indicator. In this study, the near-wall LR indicator is defined as the distance to the inner pipe wall (normalized to pipe-diameter $D$) at which the local liquid fraction is greater than the flow inlet LVF.

$$\bar{\alpha}_{lf} = \frac{\int_{c_w}^{c_w+d_s} \alpha_l \, dt}{d_s}, \qquad (10)$$

where $c_w$, $d_s$ are the coordinates of the wall and the probe sensing depth, respectively. In the study, the probe sensing depth is 5 mm.

The values of $\bar{\alpha}_{lf}$ and LR indication of water-nitrogen and oil-nitrogen flow at five different LVFs along $+x$, $-x$, $+z$, $-z$ directions along VI, MC, MD, and VO (as indicated in Fig. 3) are shown in Fig. 7a and Fig. 7b, respectively.

7(a)

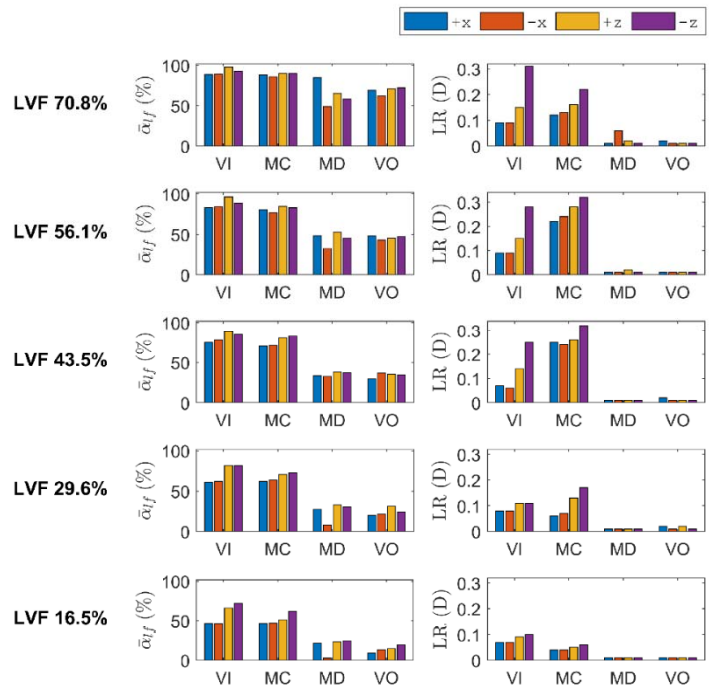

7(b)

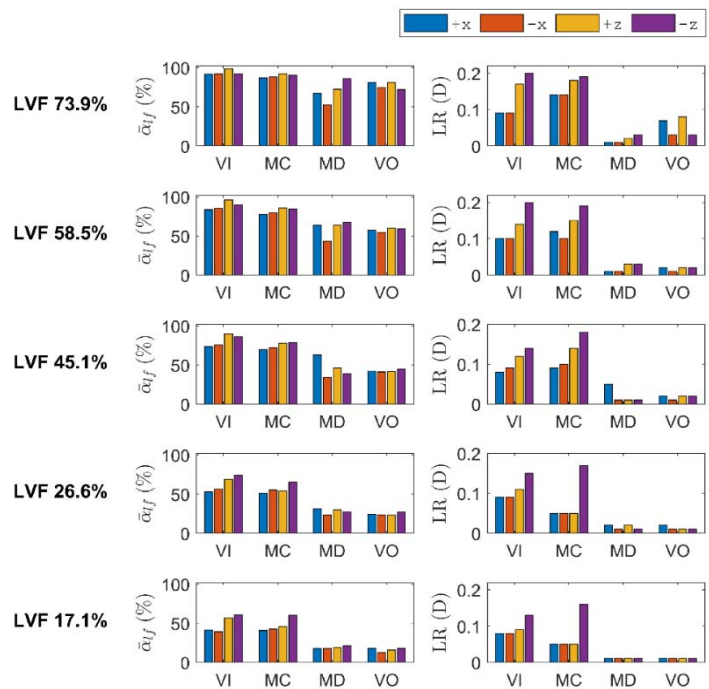

**Fig. 7.** Near-wall average liquid fraction $\bar{\alpha}_{lf}$ and liquid-richness (LR) along $+x$, $-x$, $+z$, $-z$ directions at Venturi inlet (VI), mid-convergence section (MC), mid-divergence section (MD) and Venturi outlet (VO), from (a) water-nitrogen and (b) oil-nitrogen simulations at five inlet liquid-volume-fraction (LVF) as listed.

It is observed from Fig. 7 that for both water-nitrogen and oil-nitrogen flows, $\bar{\alpha}_{lf}$ and LR at VI and MC are higher than that at MD and VO in all directions for all flow inlet LVFs tested. This suggests that VI and MC are likely to have a greater near-wall LR, which is desirable for measuring liquid property (*e.g.*, salinity) using a microwave probe. MD and VO having a low LR is probably due to liquid layer detachment and flow recirculation, as the flow entering from the section with the smaller diameter to the section with the larger diameter, as shown in Fig. 8.

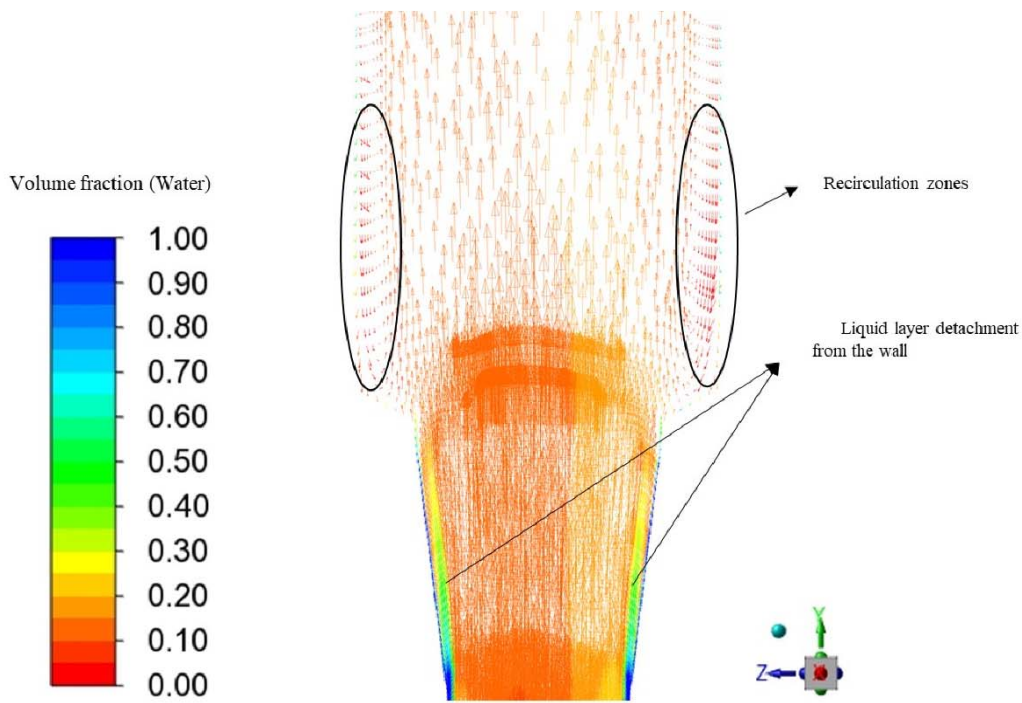

**Fig. 8.** Liquid (water) velocity vector coloured by local liquid (water) volume fraction from end-throat to Venturi outlet (water-nitrogen simulation with flow inlet liquid-volume-fraction 16.5% )

Despite having comparable LR, VI is more suitable than MC for microwave probe installation, because the latter is located between the two pressure tapping locations (located at VI and mid-throat) and used for differential pressure measurement. Locating the probe in MC may cause disturbance to the Venturi flowmeter differential pressure measurement.

In VI, the flow along the $z$ direction has higher $\bar{\alpha}_{lf}$ and LR indication than the flow along the $x$ direction, indicating greater near-wall LR along the $z$ direction. The difference in LR between the two directions increases, as the inlet LVF decreases (or as gas-volume-fraction increases).

Along the (blind-tee) $z$ direction (Fig. 3), the flow near to the $+z$ wall has a higher $\bar{\alpha}_{lf}$ than that near the $-z$ wall for the intermediate and high LVF (44% ~ 74%). At a low LVF (17% ~ 30%), the flow near to the $-z$ wall has a higher $\bar{\alpha}_{lf}$ than that near the $+z$ wall. In terms of LR indication, the flow near the $-z$ wall has a higher LR indication than the flow near the $+z$ wall, suggesting that the region where the local liquid fraction is higher than the inlet LVF extends a greater distance from the $-z$ wall than from $+z$ wall. Fig. 9 shows the water fraction distribution at VI.

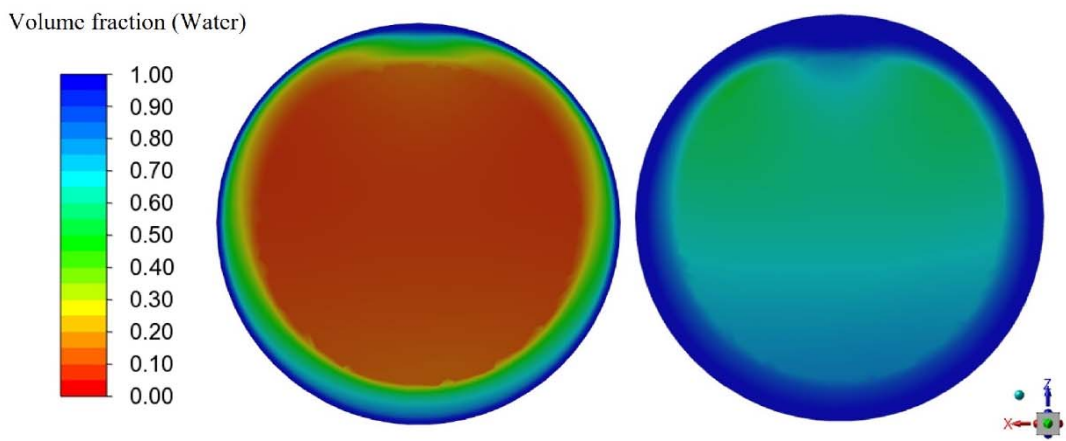

**Fig. 9.** Water volume fraction distribution at Venturi inlet (one pipe-diameter $D$ upstream of convergence section, as indicated in Fig. 3) for flow inlet liquid-volume-fraction 16.5% (left) and 70.8% (right).

It is observed that the flow in the $-z$ direction has a greater LR than that in the $+z$ direction in general, even though there is a liquid-rich spot in the $+z$ direction immediately next to the $+z$ wall. The size of the liquid-rich spot is larger in the flow with a high-inlet LVF than in that with a low-inlet LVF. This may explain why the flow with a high-inlet LVF has a higher $\bar{\alpha}_{lf}$ in the $+z$ direction than in the $-z$ direction; however, at a low-inlet LVF, the liquid-rich spot may not be large enough to extend to the probe sensing depth. Hence, at the low-inlet LVF, the flow has a higher $\bar{\alpha}_{lf}$ in the $-z$ direction than in the $+z$ direction. The flow with a low LVF is of a higher practical interest than the flow with a high LVF. The microwave probe measurement is also more challenging for a low LVF, due to the low liquid content. Hence, the $-z$ direction at VI is preferred over the $+z$ direction for the microwave probe installation.

*4.3 Experimental validation*

To validate the findings of the CFD simulation, experiments are conducted at the Schlumberger high-pressure multiphase flow test facility to assess MPFM performance. It comprises a three-phase separation system, gas compressor, two liquid pumps and a set of accurate reference flowmeters on each (oil, water and nitrogen) single-phase flow line. Pumps and compressors are used to drive the single-phase fluid coming from the separator through the corresponding reference flowmeters. After joining three single-phase flow lines, the commingled multiphase flow passes through the device under test (*e.g.*, a Venturi-based MPFM) and finally returns to the separator. The patented open-ended coaxial microwave sensor used in the experiments is engineered to withstand a pressure of 5000 psi and a temperature range of -40 to 121ºC and is $H_2S$-resistant. The pressure seal, which also acts as an electrical insulator between the inner and outer conductors, is made of appropriate glass (Xie *et al*., 2017). The glass-metal-sealed microwave probe is mounted substantially flush with the inner pipe wall in the $-z$ direction for VI, MC, MD, and VO (compared with CFD simulation therein). The other directions $+z$, $+x$ and $-x$ have the same trend as the $-z$ direction in having near-wall LR greater at VI and MC than at MD and VO. A picture of a glass-metal-sealed microwave probe and a mechanical drawing of the flush-mounted microwave probe with the inner pipe wall of the Venturi is shown in Fig. 10. The experimental setup is shown in Fig. 11.

10(a)     10(b)

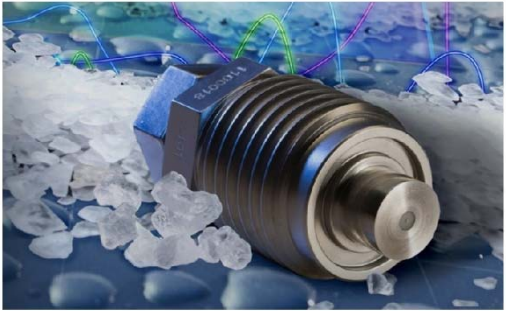
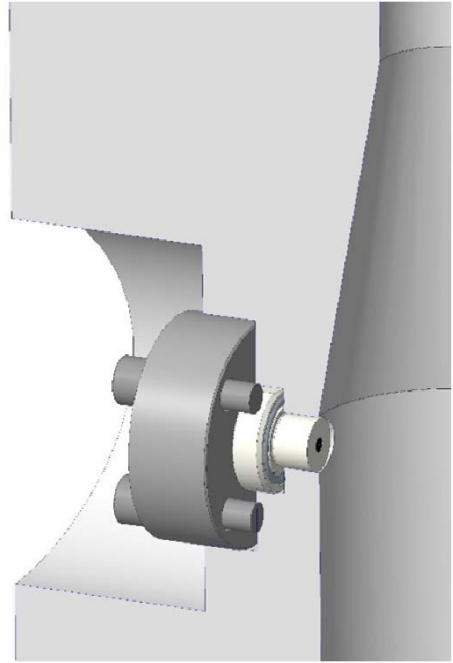

**Fig. 10.** (a) Glass-metal-sealed microwave probe used in experiments and (b) mechanical drawing showing flush-mounted microwave probe with inner pipe wall of Venturi

11(a)     11(b)

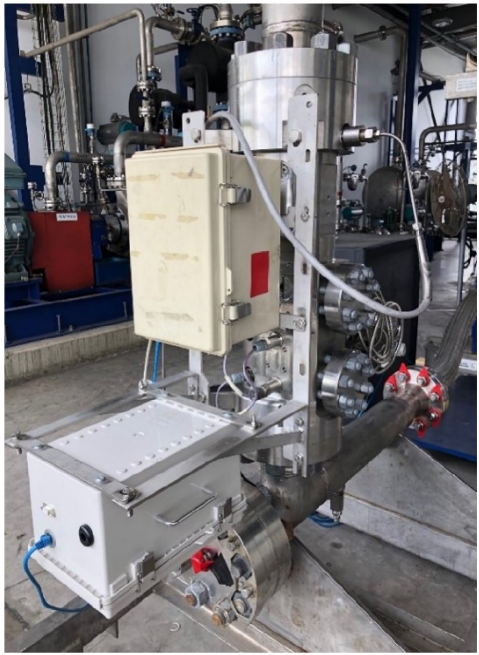
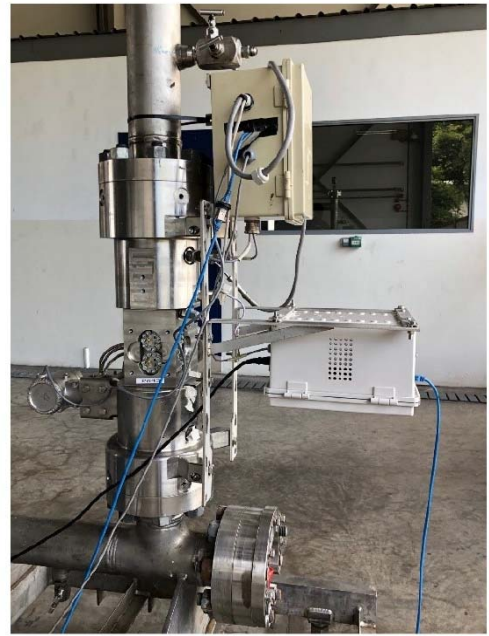

**Fig. 11.** Experimental setup for a multiphase flow loop with a glass-metal-sealed microwave probe installed along a Venturi flowmeter in (a) Venturi inlet and mid-convergence section, and (b) mid-divergence section and Venturi outlet.

For both water-nitrogen and oil-nitrogen gas-liquid flows, experimental tests are performed over a wide range of LVFs as well as gas and liquid flow rates; two test points for LVF ~ 77% and LVF ~19% are illustrated below. The near-wall apparent average liquid fraction $\bar{\alpha}_{alf}$ at the sensing depth can be computed using Equation (11) (Folgerø & Tjomsland, 1996).

$$\bar{\alpha}_{alf} = \frac{d_l}{d_s} = -\frac{C_p}{d_s}\ln\frac{\varepsilon_l - \varepsilon_m}{\varepsilon_l - \varepsilon_g}, \tag{11}$$

where $\bar{\alpha}_{alf}$ is the near-wall apparent average liquid fraction, $d_l$ is the near-wall apparent liquid layer thickness, $d_s$ is the sensing depth of the probe, $C_p$ is the empirical probe constant dependent on the probe geometry, $\varepsilon_m$ is the measured mixture dielectric permittivity, $\varepsilon_l$ is the permittivity of the liquid and $\varepsilon_g$ is the permittivity of the gas. The probe used in the study has a probe constant $C_p$ of 0.56 mm.

The normalized measured mixture permittivity $\hat{\varepsilon}_m$ (normalised to $\varepsilon_l$) and the near-wall apparent average liquid fraction $\bar{\alpha}_{alf}$ at VI, MC, MD, and VO of water-nitrogen and oil-nitrogen flows are shown in Fig. 12a and Fig. 12b, respectively.

12(a)

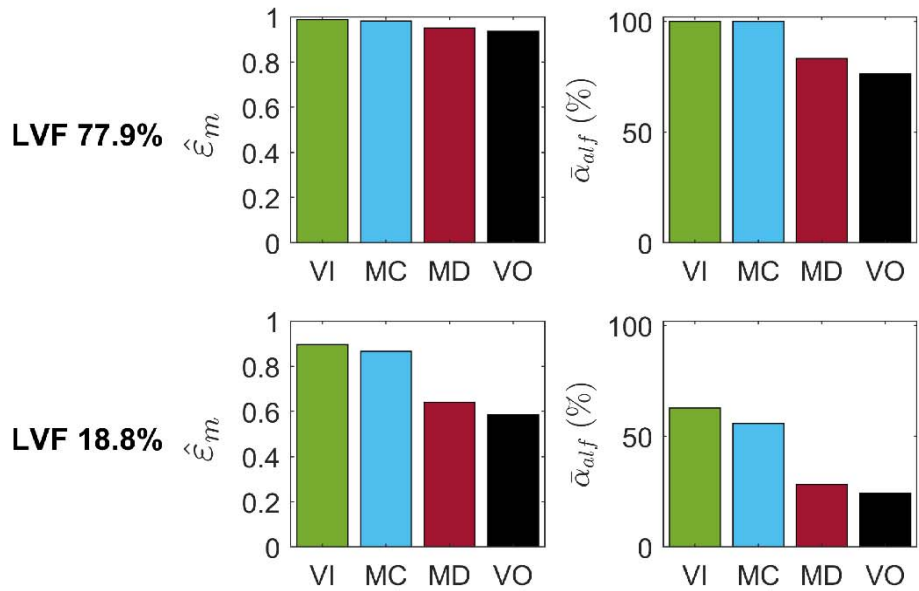

12(b)

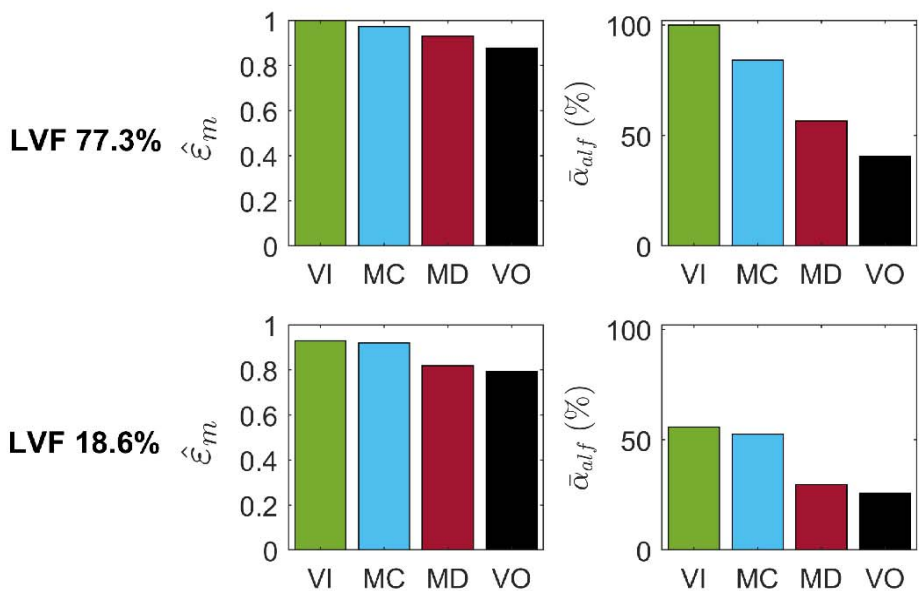

**Fig. 12.** Measured mixture permittivity $\hat{\varepsilon}_m$ (normalised to liquid permittivity $\varepsilon_l$) and near-wall apparent liquid fraction $\bar{\alpha}_{alf}$ at Venturi inlet (VI), mid-convergence section (MC), mid-divergence section (MD) and Venturi outlet (VO), liquid-volume-fraction (LVF) for (a) water-nitrogen flow ($\varepsilon_l = 71$; $\varepsilon_g = 1$) and (b) oil-nitrogen flow ($\varepsilon_l = 2.1$; $\varepsilon_g = 1$), respectively. Fluid temperature at ~ 30°C, pressure at ~ 20 bar, water salinity ~ 25 kppm.

It is observed that the normalized measured mixture permittivity $\hat{\varepsilon}_m$ and the near-wall apparent average liquid fraction $\bar{\alpha}_{alf}$ at VI and MC are higher than at MD and VO, for both illustrated

LVFs. The difference is more significant, for the flow with a low-inlet LVF. This is consistent with the CFD simulation results. Through experiments, the observation that the near-wall flow at VI and MC has greater LR than the near-wall flow at MD and VO are found to be true in a range of the Venturi throat diameter $d$ from 20 mm to 40 mm with a throat-to-inlet diameter ratio $\frac{d}{D} = 0.5$. The range of inlet LVF tested is from ~20% to ~80%. The working range of the vertically installed microwave probe at VI is experimentally found to be comparable with the existing technologies of salinity measurement at the lower area of the horizontal blind-tee. The advantage is that the entire flowmeter installation is more compact by integrating the microwave sensor with the vertically mounted MPFM. This can reduce the flowmeter manufacturing cost and the (carbon) footprint. In addition, the measurement of the flowing mixture conductivity and/or permittivity can be used to determine the liquid WLR of the gas-liquid three-phase flow. This is sometimes not feasible in horizontal blind-tee as water and oil in the horizontal blind-tee may separate due to their density difference and low local flow rates.

## 5  Conclusion

In this study, a gas-liquid two-phase CFD simulation with a steady Eulerian–Eulerian framework are used to evaluate the suitability of different microwave probe sensing locations along the vertical section of a Venturi-based flowmeter. The CFD model has been validated by the measurement data of water-nitrogen and oil-nitrogen flows in the flow loop test. Good agreement between the predicted value and the measurement is obtained in terms of the gamma-ray chord-averaged liquid fraction and the Venturi differential pressure. The one-dimensional near-wall average liquid fraction $\bar{\alpha}_{lf}$ measured over the microwave probe sensing depth and the near-wall LR is used as a criterion for assessing near-wall LR. It has been found that VI and MC have higher near-wall liquid content than MD and VO. The difference is more significant when the flow LVF is low. This is confirmed experimentally by comparing the near-wall flow mixture dielectric permittivity values measured along the Venturi by the microwave probes. VI is preferred as a microwave probe measurement location over MC. This is because MC lies between the two pressure tapping locations for the Venturi differential pressure measurement. If installed in MC, the probe may cause interference to the differential pressure measurement. The flow is more liquid-rich along (blind-tee inlet) $z$ direction than along the $x$ direction. Along the $z$ direction, the flow is generally more liquid-rich in the $-z$ direction. In terms of near-wall liquid fraction $\bar{\alpha}_{lf}$, the flow near the $+z$ wall has a higher $\bar{\alpha}_{lf}$ than that near the $-z$ wall for intermediate and high LVFs. At a low LVF, the flow near the $-z$ wall has a higher $\bar{\alpha}_{lf}$ than the flow near the $+z$ wall. Since the measurement for the wet gas flow with a low LVF is more challenging and of a high practical interest, the $-z$ wall at VI is the most suitable location for microwave probe measurement.

In summary, the contributions of this work include: (1) The microwave probe installation is integrated with a vertically mounted Venturi-based MPFM, which can potentially reduce the flowmeter (carbon) footprint without the need for a separate measurement sensor at the horizontal blind-tee. In addition to salinity measurement, WLR measurement with vertical MPFM can be improved because fluids are separated by gravity due to density difference at the horizontal blind-tee, so oil-water mixing is better at the vertical MPFM compared to horizontal blind-tee; (2) Despite the challenge of the thin near-wall liquid layer present in the vertical pipe, the VI is numerically identified and experimentally validated as a suitable location with sufficient liquid layer thickness and minimal gas entrainment for reliable microwave probe measurement among the four locations studied; (3) A systematic approach, including a CFD model that fits well with experimental measurements and post-processing methods, is developed to evaluate near-wall LR based on simulation results and experimental measurements, and could be used in future studies to identify suitable sensor locations.

**Acknowledgements**

Source: Journal of Petroleum Science and Engineering, Vol. 218, pp. 111027, 2022;
DOI: 10.1016/j.petrol.2022.111027

The authors are grateful to Chan W.L., Basman E. and Hammond P. for their constructive comments and suggestions, and Zhu J.H. and Zhan L.Y. for their help in multiphase flow loop experiments at Schlumberger and for providing the related experimental data used in this study. This work was supported by Singapore Economic Development Board and Singapore Ministry of Education Academic Research Fund Tier 1 (04MNP002133C160).